\documentclass[conference]{IEEEtran}

\newtheorem{theorem}{Theorem}
\newtheorem{lemma}{Lemma}
\newtheorem{corollary}{Corollary}
\newtheorem{definition}{Definition}
\newtheorem{remark}{Remark}
\newtheorem{proposition}{Proposition}
\usepackage{multirow}

\usepackage[cmex10]{amsmath}
\usepackage{amssymb}

\usepackage[dvips]{graphicx}
\usepackage{psfrag}

\IEEEoverridecommandlockouts
\begin{document}
%%%%%%%%%%%%%%%%%%%%%%%%%%%%%%%%%%%%%%%%%%%%%%%%%%%%%______Title______%%%%%%%%%%%%%%%%%%%%%%%%%%%%%%%%%%%%%%%%%%%%%%%%%%%%%%%

\title{Completion Time in Multi-Access Channel: An Information Theoretic Perspective}
\author{\IEEEauthorblockN{Yuanpeng Liu, Elza Erkip}
\IEEEauthorblockA{ECE Department, Polytechnic Institute of New York University\\
yliu20@students.poly.edu, elza@poly.edu}
\thanks{This work was partially supported by NSF grant No. 0635177.}}
\maketitle

%%%%%%%%%%%%%%%%%%%%%%%%%%%%%%%%%%%%%%%%%%%%%%%%%%%%%____Abstract_____%%%%%%%%%%%%%%%%%%%%%%%%%%%%%%%%%%%%%%%%%%%%%%%%%%%%%%%
\begin{abstract}
In a multi-access channel, completion time refers to the number of channel uses required for users, each with some given fixed bit pool, to complete the transmission of all their data bits. In this paper, the characterization of the completion time region is based on the concept of constrained rates, where users' rates are defined over possibly different number of channel uses. An information theoretic formulation of completion time is given and the completion time region is then established for two-user Gaussian multi-access channel, which, analogous to capacity region, characterizes all possible trade-offs between users' completion times.

\end{abstract}

%%%%%%%%%%%%%%%%%%%%%%%%%%%%%%%%%%%%%%%%%%%%%%%%%%%____Introduction_____%%%%%%%%%%%%%%%%%%%%%%%%%%%%%%%%%%%%%%%%%%%%%%%%%%%%%%%
\section{Introduction}
Multi-access channel (MAC) is an important channel model that finds many applications in wireless networks and has drawn substantial research attention in the literature. Traditionally the study of MAC is guided under two different philosophies. Information theoretic approach assumes users have a full buffer all the time and strives to characterize the fundamental limits of transmission rates by focusing on the interplay of noise and interference \cite{Ahlswede}\cite{Liao}. In contrast, network oriented studies usually focus on addressing issues arising from bursty packet arrivals and view the multi-user channel as a collection of single user channels by adopting a collision model \cite{Abramson}, treating interference as noise or orthogonalization in time or frequency domain. The discrepancy between these two lines of work has been well documented in \cite{Gallager}\cite{Ephremides}. In this paper, we use information theoretic tools to study a problem, which deviates from the usual information theoretic setup and has a flavor of network theory, namely the completion time in multi-access channel. The completion time problem attempts to incorporate the notion of delay into the information theoretic study of MAC.

This paper considers a periodic source arrival model, where a new block of data arrives very $n$ channel uses. Hence during each block of $n$ channel uses, user's data buffer is not to be replenished and the usual full-buffer assumption is no longer valid. One example of this would be users sending large files of fixed sizes, in which case we only have a single channel block. For another example, consider two users streaming live videos that are compressed at possibly different but fixed rates to a common receiver. The data arrives periodically. In the beginning of each period, each user has a certain number of bits to send. However due to the casuality constraint, after the completion of the current transmission, users will have to wait until the next period to obtain new data to send. We model this as the follows: user $i$, $i=1,2$, has $m\tau_i$ bits, with $m\tau_i$ corresponding to the file size in the file example and $\tau_i$ corresponding to the compression rate in the streaming example, to transmit in at most $n$ channel uses, where $n$ is assumed to be large enough to allow the completion of both transmissions. Let $n_i \leq n$ be the actual number of channel uses that user $i$ spends on the transmission. We are interested in the \textit{normalized completion time} (hereafter referred as ``completion time'') within a single channel block, which is defined as $n_i/m$ in the limit of large $n_i$ and $m$. Note that in the streaming example, $m$ corresponds to the number of source samples, which is assumed to be the same for both users. In general we can view $m$ as a scaling factor to ensure information theoretic arguments with large block lengths can be invoked. The exact value of $m$ is not important since it will not appear in the characterization of completion time.

The main contributions of this paper are an information theoretic formulation of completion time and the derivation of the completion time region for two-user Gaussian multi-access channel (GMAC), which, analogous to capacity region, characterizes all possible trade-offs between users' completion times. Compared with \cite{Rai}, where the authors solved the sum completion time minimization problem for a $K$-user symmetric GMAC, our result provides a more general formulation for the two-user case. In \cite{Ng}, the authors considered an interference channel where each user has backlogged packets of equal size to transmit and the goal is to leverage power control to minimize some convex cost function over the completion time region. In \cite{Ng}, the completion time region is obtained by treating interference as noise, whereas in this paper, we adopt an information theoretic approach without restricting ourselves to any specific coding scheme such as treating interference as noise.

This paper is organized as the follows. In Section II, the concept of constrained rates is introduced, based on which completion time is then defined. In Section III, the completion time region for two-user GMAC is derived. Applications of the obtained completion time region and extensions of this work are discussed in Section IV.

\emph{Notation}: Let $\gamma(x)=\frac{1}{2}\textrm{log}_2(1+x)$. Also let $X_{k,i}^j=(X_{k,i},...,X_{k,j})$ for $i\leq j$ and $X_k^j=X_{k,1}^j=(X_{k,1},...,X_{k,j})$. $X_{k,i}^j$ does not appear if $i>j$. $[X]^+=\max\{X,0\}$. We use bold font for vectors and calligraphic font for regions.

%%%%%%%%%%%%%%%%%%%%%%%%%%%%%%%%%%%%%%%%%%%%%%%%%%%____Formulation_____%%%%%%%%%%%%%%%%%%%%%%%%%%%%%%%%%%%%%%%%%%%%%%%%%%%%%%%
\section{Problem Formulation}
In information theory, it is commonly assumed that a user's data buffer is always full and therefore the goal is to devise a coding scheme that can reliably transmit as much information as possible in a given number of channel uses. However, in our setup, the total amount of information to be transmitted for user $i$ is limited to $m\tau_i$ bits in at most $n$ channel uses. Hence having some users finish early is not only desirable to reduce to completion time of those users, but also preferable for the remaining users since they can enjoy reduced multi-user interference in the remaining period. In order to capture this, and to formulate the completion time problem, we will define communication rates over different number of channel uses for each user, as opposed to the standard definition in multi-user information theory, where users' codewords span the same block length. We refer this as {\em constrained rate}, which will be first defined through a two-user discrete memoryless multi-access channel (DMMAC) in subsection II.A. We then give a formal definition of completion time in subsection II.B.

%_____________________________________________________________________________________________________________________________
\subsection{Constrained Rate}
Consider a two-user DMMAC $(\mathcal{X}_1\times\mathcal{X}_2,p(y|x_1,x_2),\mathcal{Y})$, where $\mathcal{X}_1,\mathcal{X}_2$ are the input alphabets, $\mathcal{Y}$ is the channel output alphabet and $p(y|x_1,x_2)$ is the channel transition probability.

Let us denote $n=\max\{n_i\}$, $i=1,2$, and $c=n_1/n_2$. We will let $n_1$ and $n_2$ vary with $c$ fixed. For $i=1,2$, $\bar{i}=\{1,2\}\setminus i$. Also let
\begin{equation}
    R_i^*=\max_{p_{X_i}} I(X_i;Y|X_{\bar{i}}=\phi_{\bar{i}}), \quad i=1,2,\label{ratestar}
\end{equation}
where $\phi_{\bar{i}}=\arg \max_{\phi\in\mathcal{X}_{\bar{i}}}\max_{p_{X_i}} I(X_i;Y|X_{\bar{i}}=\phi)$. One can view $\phi_{\bar{i}}$ as the symbol that ``opens'' up the channel from user $i$ to the receiver the most.

\begin{definition}
A $((M_1,M_2),n_1,n_2)$ code consists of message sets: $\mathcal{W}_i=\{1,...,M_i\}$, two encoding functions,
\begin{align*}
    X_i:\mathcal{W}_i\rightarrow (\mathcal{X}_i^{n_i},\phi_{i,{n_i+1}}^n)\quad \textrm{for } i=1,2
\end{align*}
and two decoding functions
\begin{align*}
    g_i:\mathcal{Y}^{n_i}\rightarrow \mathcal{W}_i\quad \textrm{for } i=1,2
\end{align*}
Note that user $i$ will send $\phi_i$ during the $n-n_i$ symbols at the end of its codeword.
\end{definition}

Users independently choose an index $W_i$ uniformly from $\mathcal{W}_i$ and send the corresponding codewords. The average error probability for the $((M_1,M_2),n_1,n_2)$ code is
\begin{equation*}
    P_e=\textrm{Pr}(g_1(Y^{n_1})\neq W_1 \textrm{ or } \ g_2(Y^{n_2})\neq W_2).
\end{equation*}
\begin{definition}
For a $((M_1,M_2),n_1,n_2)$ code, the \textit{$c$-constrained rates} are defined as, for $i=1,2$,
\begin{equation}
\label{ratedef}
    R_i=\frac{\log_2(M_i)}{n_i} \quad \textrm{bits per channel use}.
\end{equation}
\end{definition}

The $c$-constrained rate pair $(R_1,R_2)$ is said to be \textit{achievable} if there exists a sequence of $((M_1,M_2),n_1,n_2)$ codes with $P_e\rightarrow 0$ as $n\rightarrow \infty$. The \textit{c-constrained rate region}, denoted by $\mathcal{R}_c$, is the set of achievable $c$-constrained rate pairs for a given coding scheme. The \textit{c-constrained capacity region} $\mathcal{C}_c$ is the closure of all $\mathcal{R}_c$.
\begin{remark}
We use the term ``$c$-constrained rate (capacity) region'' to emphasize the fact that user $i$'s effective codeword length is constrained by $n_i$ channel uses over which $R_i$ is defined (the remaining channel uses are padded by $\phi_i$) and the rate (capacity) region is a function of $c=n_1/n_2$. Hence $\mathcal{R}_1$ ($\mathcal{C}_1$) is the standard rate (capacity) region, where $n_1=n_2$. Lemma \ref{raterelation} and Theorem \ref{maccapacity}, which will be stated next, reveal the connection between the $c$-constrained rate (capacity) region and the standard one.
\end{remark}
\begin{lemma}
\label{raterelation}
The $c$-constrained rate pair $(R_1,R_2)$ is achievable, for some $c\neq 1$, if:
\begin{enumerate}
\item{$c<1$,}
$R_2$ can be decomposed into $R_2'$ and $R_2''$: $R_2=cR_2'+(1-c)R_2''$, such that $(R_1,R_2')\in\mathcal{C}_1$, $R_2''\leq R_2^*$;

\item{$c>1$,}
$R_1$ can be decomposed into $R_1'$ and $R_1''$: $R_1=\frac{1}{c}R_1'+(1-\frac{1}{c})R_1''$, such that $(R_1',R_2)\in\mathcal{C}_1$, $R_1''\leq R_1^*$,

\end{enumerate}
where $R_i^*$ is defined in (\ref{ratestar}).
\end{lemma}
\begin{IEEEproof}
The proof is relegated to Appendix \ref{Proofraterelation}.
\end{IEEEproof}
\begin{theorem}
\label{maccapacity}
The $c$-constrained capacity region $\mathcal{C}_c$ for some $c\neq 1$ is the set of all rate pairs $(R_1,R_2)$ satisfying
\begin{enumerate}
\item{$c<1$, $(R_1,[\tfrac{1}{c}R_2-(\tfrac{1}{c}-1)R_2^*]^+)\in\mathcal{C}_1$;}
\item{$c>1$, $([cR_1-(c-1)R_1^*]^+,R_2)\in\mathcal{C}_1$,}
\end{enumerate}
where $\mathcal{C}_1$, the standard two-user DMMAC capacity region, is the closure of the set of all $(r_1,r_2)$ pairs satisfying
\begin{align*}
    r_1&\leq I(X_1;Y|X_2,Q),\\
    r_2&\leq I(X_2;Y|X_1,Q),\\
    r_1+r_2&\leq I(X_1,X_2;Y|Q)
\end{align*}
for some choice of the joint distribution $p(q)p(x_1|q)p(x_2|q)p(y|x_1,x_2)$ with $|Q|\leq 4$. To avoid confusion, hereafter we use lower-case $r$ and upper-case $R$ to refer to the standard and constrained rates respectively.
\end{theorem}
\begin{IEEEproof}
The proof is relegated to Appendix \ref{Proofmaccapacity}.
\end{IEEEproof}

Now let us consider the following two-user GMAC:
\begin{align}
    Y=X_1+X_2+Z,\label{GMACchannel}
\end{align}
where $Z\sim\mathcal{N}(0,1)$ is the i.i.d. Gaussian noise process and inputs are subject to per symbol power constraints: $E[X_i^2]\leq P_i$. The standard capacity region of two-user GMAC is
\begin{align}
    \label{GMACconvcap}
    \mathcal{C}_1^G=\left\{(r_1,r_2)\Big | 0\leq \sum_{i\in \Omega}r_i\leq \gamma(\sum_{i\in \Omega}P_i),\ \Omega\subseteq \{1,2\}\right\}.
\end{align}
To avoid confusion, hereafter we use lower-case $r$ and upper-case $R$ to refer to the standard and constrained rates respectively. Note that for the GMAC, $\phi_i=0$, i.e. user $i$ stays silent after it completes the transmission in $n_i$ channel uses.

\begin{corollary}
\label{GMACcapacity}
The $c$-constrained capacity region $\mathcal{C}_c^G$ of two-user GMAC is the set of non-negative rate pairs $(R_1,R_2)$ satisfying:
\begin{align*}
    R_1\leq \gamma(P_1),&\quad R_2\leq \gamma(P_2),\\
    \max(1,c)R_1+\max(1,\tfrac{1}{c})R_2&\leq (c-1)\gamma(P_1)I\\
    &+(\tfrac{1}{c}-1)\gamma(P_2)\bar{I} +\gamma(P_1+P_2),
\end{align*}
where $\bar{I}=1-I$ and $I$ is an index function: $I=1$ if $c\geq 1$; $I=0$ if $c< 1$.
\begin{IEEEproof}
The proof mostly follows from Theorem \ref{maccapacity}. The achievability is obtained by using Gaussian input distribution. Substituting $\mathcal{C}_1^G$ for $\mathcal{C}_1$ and $\gamma(P_i)$ for $R_i^*$ in Theorem \ref{maccapacity}, the above expression can be derived after some manipulation. The converse follows from the converse of Theorem \ref{maccapacity} and the fact that (\ref{macr15}), (\ref{macr22}), (\ref{macrss}) and $R_i^*$ are maximized by Gaussian input distribution under the per symbol power constraints: $E[X_i^2]\leq P_i$.
\end{IEEEproof}
\end{corollary}

%______________________________________________________________________________________________________________________________
\subsection{The Notion of Completion Time}
Consider a two-user DMMAC, where each user has $m\tau_i$ ($i=1,2$) bits to send to a common receiver.

\begin{definition}
\label{delaydef}
We define the \textit{normalized completion time} as $d_i=n_i/m$, where $n_i$ is the actual number of channel uses that user $i$ spends on transmitting $m\tau_i$ bits.
\end{definition}

Because of the relation $\log_2(M_i)=n_iR_i=m\tau_i$ in (\ref{ratedef}), where $R_i$ is the $c$-constrained rate, we have $d_i=\tau_i/R_i$. Completion time pair $(d_1,d_2)$ is said to be \textit{achievable} if $(\tau_1/d_1,\tau_2/d_2)$ is an achievable $c$-constrained rate pair, i.e. $(\tau_1/d_1,\tau_2/d_2)\in \mathcal{R}_c$ where $c=n_1/n_2=d_1/d_2$. The achievable completion time region for a given coding scheme is $\mathcal{D}=\{(d_1,d_2)|(\tau_1/d_1,\tau_2/d_2)\in \mathcal{R}_{d_1/d_2}\}$. Analogous to capacity region, we can also define the overall completion time region $\mathcal{D}^*$ as the union of all achievable completion time regions, or equivalently $\mathcal{D}^*=\{(d_1,d_2)|(\tau_1/d_1,\tau_2/d_2)\in \mathcal{C}_{d_1/d_2}\}$.

%%%%%%%%%%%%%%%%%%%%%%%%%%%%%%%%%%%%%%%%%%%%%%%____Completion Time Region_____%%%%%%%%%%%%%%%%%%%%%%%%%%%%%%%%%%%%%%%%%%%%%%%%%%%%
\section{Completion Time Region for Two-user GMAC}
In this section, we establish properties of $\mathcal{D}^*$ and compute it for two-user GMAC. Notice that an achievable completion time pair $(d_1,d_2)$ is defined in terms of $c$-constrained rate pair, which in return depends on $(d_1,d_2)$ through $c=d_1/d_2$. Hence it is easy to check for a given $(d_1,d_2)$ whether or not it is achievable, but difficult to directly compute all pairs of $(d_1,d_2)\in\mathcal{D}^*$ using the definition because of this recursive dependence. Another difficulty in determining $\mathcal{D}^*$ is that it is not always convex, as we shall show later in Theorem \ref{CTR}. To walk around these obstacles, we characterize two sub-regions of $\mathcal{D}^*$ seperately and the union of the two will lead us to $\mathcal{D}^*$. In subsection III.A, we first show that the sub-regions are always convex. In subsection III.B, we consider the weighted sum completion time minimization problem over the sub-regions for two-user GMAC, which will be used in subsection III.C to show the achievability and converse when we establish the sub-regions and hence $\mathcal{D}^*$.

%______________________________________________________________________________________________________________________________
\subsection{Convexity of Sub-regions of $\mathcal{D}^*$}
\begin{proposition}
\label{convex}
$\mathcal{D}^*$ contains two convex sub-regions, $\mathcal{D}_1^*$ and $\mathcal{D}_2^*$, where
\begin{align*}
    \mathcal{D}_{1}^*&=\mathcal{D}^*\bigcap\{(d_1,d_2)|d_1\leq d_2\},\\
    \mathcal{D}_{2}^*&=\mathcal{D}^*\bigcap\{(d_1,d_2)|d_1\geq d_2\}.
\end{align*}
\end{proposition}
\begin{IEEEproof}
We prove for $\mathcal{D}_{1}^*$, i.e. $d_1\leq d_2$. The case of $\mathcal{D}_{2}^*$ follows similarly. For $\mathbf{d},\mathbf{d}'\in\mathcal{D}_{1}^*$, we need to show $\mathbf{d}''=\alpha\mathbf{d}+\bar{\alpha}\mathbf{d}'\in\mathcal{D}_{1}^*$, where $0\leq\alpha\leq 1$ and $\bar{\alpha}=1-\alpha$.

Suppose $\mathbf{d}=(d_1,d_2)$ is an achievable completion time pair. We consider the transmission scheme pictorially depicted in Fig \ref{timesharing}.a. In the first $n_1=md_1$ channel uses, the two users employ some coding scheme, denoted by $SCH_1$, where proper decoding is required at the end of $n_1$ channel uses to ensure messages sent in this time interval are received correctly. In the remaining $n_2-n_1=m(d_2-d_1)$ channel uses, coding scheme $SCH_2$ is employed at user 2 while user 1 sends $\phi_1$. The decoding for this part of user 2's message is done at the end of $n_2-n_1$ channel uses. Note that by Theorem \ref{maccapacity} and Lemma \ref{raterelation}, we can view user 2's message consisting of two independent parts for $n_1$ and $n_2-n_1$ channel uses respectively. Similarly for $\mathbf{d}'$, we consider coding schemes $SCH_1'$ and $SCH_2'$ shown in Fig \ref{timesharing}.b. Based on the coding schemes for $\mathbf{d}$ and $\mathbf{d}'$, we construct a new coding scheme, depicted in Fig \ref{timesharing}.c. Since during each sub-interval error probability can be driven arbitrarily small, the overall scheme is reliable. The completion time achieved by this scheme for user $i$ ($i=1,2$) is $\frac{\alpha n_i+\bar{\alpha}n_i'}{m}=\alpha d_i+\bar{\alpha}d_i'$. Furthermore, since $d_1\leq  d_2$ and $d_1'\leq d_2'$, we have $\alpha d_1+\bar{\alpha}d_1'\leq\alpha d_2+\bar{\alpha}d_2'$. Therefore $\mathbf{d}''\in\mathcal{D}_1^*$.
\begin{figure}[htb]
    \centering
    \includegraphics[width=65mm]{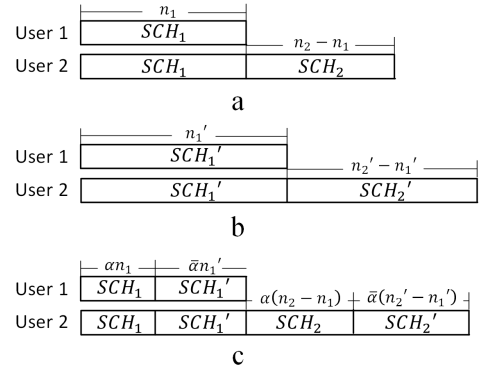}
    \caption{Sub-figures a, b and c depict the transmission schemes that achieve $\mathbf{d}$, $\mathbf{d}'$ and $\mathbf{d}''$ respectively.}
    \label{timesharing}
\end{figure}
\end{IEEEproof}
\begin{remark}
The proof of Proposition \ref{convex} relies on the condition that for $\mathbf{d}$, $\mathbf{d}'$ if $d_1\leq d_2$, then $d_1'\leq d_2'$, vice versa. Without this condition, it is impossible to derive a simple transmission scheme, based on the coding schemes for $\mathbf{d}$ and $\mathbf{d}'$ respectively, that achieves the convex combined completion time $\mathbf{d}''$. This is because, in this case, the codewords in different sub-intervals will not be aligned, as opposed to the case shown in Fig \ref{timesharing}.c., and thus we cannot argue that the decoding in each sub-interval (hence for the overall transmission scheme) will be successful.
\end{remark}

%______________________________________________________________________________________________________________________________
\subsection{Weighted Sum Completion Time Minimization Problem}
For the remainder of this paper, we will focus on GMAC. In this subsection, we solve the following weighted sum completion time minimization problem:
\begin{align}
    \textrm{minimize    }&\quad d_s=wd_1+\bar{w}d_2 \label{optorigin}\\
    \textrm{subject to  }&\quad (d_1,d_2)\in\mathcal{D}_{i}^*,\ i=1,2 \notag
\end{align}
where $\bar{w}=1-w$ and $w\in[0,1]$. We first transform (\ref{optorigin}) into an equivalent problem using the connection between the $c$-constrained capacity region and the standard capacity region described in section II.A. We begin by introducing some notations. Define
\begin{align}
\label{Di}
    D_1=\bar{w}\tfrac{\tau_2}{R_2^*}+\tfrac{\tau_1(R_2^*-\bar{w}r_2)}{R_2^*r_1}, \ D_2=w\tfrac{\tau_1}{R_1^*}+\tfrac{\tau_2(R_1^*-wr_1)}{R_1^*r_2},
\end{align}
where $R_i^*=\gamma(P_i)$, i.e. (\ref{ratestar}) evaluated for Gaussian channel (\ref{GMACchannel}). We use $D_i(\mathbf{r})$ to denote the value of $D_i$ evaluated at $\mathbf{r}=(r_1,r_2)$. We also define
\begin{align*}
    \mathcal{C}_{1,1}^G=\mathcal{C}_1^G\bigcap\{(r_1,r_2)|\tfrac{r_2}{r_1}\leq \tfrac{\tau_2}{\tau_1}\},\\
    \mathcal{C}_{1,2}^G=\mathcal{C}_1^G\bigcap\{(r_1,r_2)|\tfrac{r_2}{r_1}\geq \tfrac{\tau_2}{\tau_1}\},
\end{align*}
where $\mathcal{C}_1^G$ is given by (\ref{GMACconvcap}).
\begin{proposition}
\label{eq}
The following optimization problem is equivalent to (\ref{optorigin}):
\begin{align}
    \textrm{minimize    }&\quad D_i \label{newopt}\\
    \textrm{subject to  }&\quad (r_1,r_2)\in\mathcal{C}_{1,i}^G, \ i=1,2\notag
\end{align}
We use $D_i^*$ to denote the optimal value in (\ref{newopt}).
\end{proposition}
\begin{IEEEproof}
Let's first consider (\ref{optorigin}) with $i=1$, i.e. $d_1\leq d_2$ according to the definition of $\mathcal{D}_1^*$ in Proposition \ref{convex}. $(d_1,d_2)\in\mathcal{D}_1^*$ implies, by definition, $(\tau_1/d_1,\tau_2/d_2)\in\mathcal{C}^G_c$, where $c=d_1/d_2\leq 1$. Without loss of generality, consider $R_2''=R_2^*$, i.e. letting user 2 transmit at the maximum point to point rate in the second phase in order to minimize its delay. According to Theorem \ref{maccapacity}, for some $(r_1,r_2)\in\mathcal{C}^G_1$ , we have $r_1=R_1$ and $r_2=\frac{1}{c}R_2-(\frac{1}{c}-1)R_2^*$, where $(R_1,R_2)$ is an achievable $c$-constrained rate pair. Substituting $R_i=\tau_i/d_i$ and $c=d_1/d_2$, we have the following relations:
\begin{align}
\label{d1}
    d_1=\tfrac{\tau_1}{r_1},\quad d_2=\tfrac{\tau_2}{R_2^*}+\tfrac{(R_2^*-r_2)\tau_1}{R_2^*r_1}.
\end{align}
Hence $d_s=wd_1+\bar{w}d_2=D_1$ for $c\leq 1$. Furthermore $d_1\leq d_2$ reduces to $\frac{r_2}{r_1}\leq \frac{\tau_2}{\tau_1}$. Following the same steps, we can show
\begin{align}
\label{d2}%%%
    d_1=\tfrac{\tau_1}{R_1^*}+\tfrac{(R_1^*-r_1)\tau_2}{R_1^*r_2},\quad d_2=\tfrac{\tau_2}{r_2}.
\end{align}
and $d_s=D_2$ for $c\geq 1$. Also $d_1\geq d_2$ reduces to $\frac{r_2}{r_1}\geq \frac{\tau_2}{\tau_1}$. Therefore, the optimization problem (\ref{newopt}) is equivalent to (\ref{optorigin}).
\end{IEEEproof}

\begin{figure}[htb]
    \centering
    \includegraphics[width=65mm]{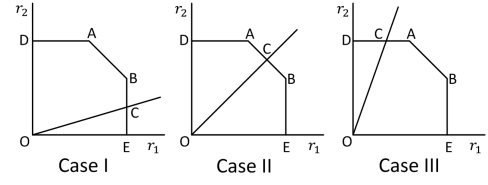}
    \caption{There are three cases of the position of point C, the intersection of the line $r_2/r_1=\tau_2/\tau_1$ and $\mathcal{C}_1^G$.}
    \label{MAC capacity region}
\end{figure}

Before we solve (\ref{newopt}), we introduce some more notations. Let $A$ and $B$, shown in Fig \ref{MAC capacity region}, denote the two corner points of $\mathcal{C}_1^G$ in (\ref{GMACconvcap}), where $A=(\gamma(P_1+P_2)-\gamma(P_2), \gamma(P_2))$ and $B=(\gamma(P_1),\gamma(P_1+P_2)-\gamma(P_1))$. Let point $C$ denote the intersection of the line $r_2/r_1=\tau_2/\tau_1$ and $C_1^G$. We define three cases depending on the position of point $C$:
\begin{enumerate}
\item Case I: $\frac{\tau_2}{\tau_1}\leq \frac{\gamma(P_1+P_2)-\gamma(P_1)}{\gamma(P_1)}$, $C=(\gamma(P_1),\tfrac{\tau_2}{\tau_1}\gamma(P_1))$;

\item Case II: $\frac{\gamma(P_1+P_2)-\gamma(P_1)}{\gamma(P_1)}<\frac{\tau_2}{\tau_1}
    <\frac{\gamma(P_2)}{\gamma(P_1+P_2)-\gamma(P_2)}$,

$\qquad \quad \ \ C=(\tfrac{\tau_1}{\tau_1+\tau_2}\gamma(P_1+P_2),\tfrac{\tau_2}{\tau_1+\tau_2}\gamma(P_1+P_2))$;

\item Case III: $\frac{\tau_2}{\tau_1}\geq \frac{\gamma(P_2)}{\gamma(P_1+P_2)-\gamma(P_2)}$, $C=(\tfrac{\tau_1}{\tau_2}\gamma(P_2),\gamma(P_2))$.

\end{enumerate}

One can think of equations (\ref{d1}) and (\ref{d2}) as functions that map a rate pair $(r_1,r_2)$ to a completion time pair depending on whether $d_1\leq d_2$ or $d_1\geq d_2$. Hence we use $\mathbf{d}_i(\mathbf{r})$ to denote the completion time pair $(d_1,d_2)$ evaluated at rate $\mathbf{r}=(r_1,r_2)$, where (\ref{d1}) is used if $i=1$ and (\ref{d2}) is used if $i=2$.

\begin{theorem}
\label{mini}
The solution to the optimization problem (\ref{optorigin}) is summarized in Table \ref{table}, where $w_1= \tfrac{\gamma(P_1+P_2)-\gamma(P_2)}{\gamma(P_1+P_2)}$ and $w_2=\tfrac{\gamma(P_1)}{\gamma(P_1+P_2)}$.
\begin{center}
\begin{table}[htbp]
\caption{}
\label{table}
\hfill{}
\begin{tabular}{|c|c|c|c|c|}
\hline
                       &                &     Case I        &      Case II       &    Case III \\
\hline
\multirow{2}{*}{$i=1$} &  $w\in[0,w_1]$ & $\mathbf{d}_1(C)$ & $\mathbf{d}_1(C)$  & $\mathbf{d}_1(A)$\\\cline{2-5}
                       &  $w\in(w_1,1]$ & $\mathbf{d}_1(C)$ & $\mathbf{d}_1(B)$  & $\mathbf{d}_1(B)$\\\hline
\multirow{2}{*}{$i=2$} &  $w\in[0,w_2]$ & $\mathbf{d}_2(A)$ & $\mathbf{d}_2(A)$  & $\mathbf{d}_2(C)$\\\cline{2-5}
                       &  $w\in(w_2,1]$ & $\mathbf{d}_2(B)$ & $\mathbf{d}_2(C)$  & $\mathbf{d}_2(C)$\\
\hline
\end{tabular}
\hfill{}
\end{table}
\end{center}
\end{theorem}
\begin{IEEEproof}
The proof is relegated to Appendix \ref{Proofmini}.
\end{IEEEproof}

%______________________________________________________________________________________________________________________________
\subsection{Completion Time Region}
Before presenting the main result on the completion time region, we also provide an outer-bound of $\mathcal{D}^*$. According to Corollary \ref{GMACcapacity}, the constrained rate $R_i$, $i=1,2$, is upper-bounded by the point to point rate $\gamma(P_i)$. Hence user $i$'s completion time is lower-bounded by $\tau_i/\gamma(P_i)$.
\begin{lemma}
\label{openregion}
$\mathcal{D}^*$ is outer-bounded by $\mathcal{D}^o$, i.e. $\mathcal{D}^*\subseteq \mathcal{D}^o$, where $\mathcal{D}^o= \{(d_1,d_2)|d_1\geq \tau_1/\gamma(P_1),d_2\geq \tau_2/\gamma(P_2)\}$.
\end{lemma}

\begin{figure}[htb]
    \centering
    \includegraphics[width=90mm]{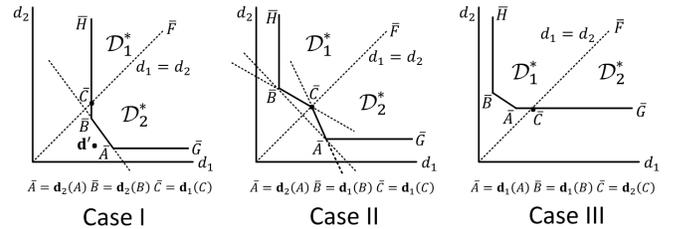}
    \caption{The completion time region $\mathcal{D}^*=\mathcal{D}_1^*\bigcup \mathcal{D}_2^*$.}
    \label{CTRegion}
\end{figure}
\begin{theorem}
\label{CTR}
The completion time region $\mathcal{D}^*$ of two-user GMAC (illustrated in Fig. \ref{CTRegion}) is the set of pairs $(d_1,d_2)$ satisfying the following:
\begin{enumerate}
\item{Case I}
\begin{align*}
    &\gamma(P_1)d_1\geq \tau_1,\ \gamma(P_2)d_2\geq \tau_2,\\
    &\gamma(P_1)d_1+[\gamma(P_1+P_2)-\gamma(P_1)]d_2\geq \tau_1+\tau_2.
\end{align*}
\item{Case II}
\begin{align*}
    &\gamma(P_1)d_1\geq \tau_1,\ \gamma(P_2)d_2\geq \tau_2,\\
    &[\gamma(P_1+P_2)-\gamma(P_2)]d_1+\gamma(P_2)d_2\geq \tau_1+\tau_2,\\
    &\gamma(P_1)d_1+[\gamma(P_1+P_2)-\gamma(P_1)]d_2\geq \tau_1+\tau_2.
\end{align*}
\item{Case III}
\begin{align*}
    &\gamma(P_1)d_1\geq \tau_1,\ \gamma(P_2)d_2\geq \tau_2,\\
    &[\gamma(P_1+P_2)-\gamma(P_2)]d_1+\gamma(P_2)d_2\geq \tau_1+\tau_2.
\end{align*}
\end{enumerate}
\end{theorem}

\begin{IEEEproof}
We will first characterize $\mathcal{D}_1^*$ and $\mathcal{D}_2^*$ separately and the union of the two gives us $\mathcal{D}^*$. We prove Case I. The others follow similarly.

We first prove $\mathcal{D}_1^*=\{(d_1,d_2)|d_1\geq \tau_1/\gamma(P_1), d_1\leq d_2\}$. For the achievability, notice that, by Theorem \ref{mini}, $\bar{C}=\mathbf{d}_1(C)=(\tau_1/\gamma(P_1),\tau_1/\gamma(P_1))$ solves (\ref{optorigin}). Thus $\bar{C}\in\mathcal{D}_1^*$, i.e. $\bar{C}$ is achievable. Referring to Fig. \ref{CTRegion} Case I, any point on the ray $\bar{C}\bar{H}$ is achievable. This is because we can use the same codebooks designed for achieving $\bar{C}$ but decrease the rate of user 2 by only using part of the codewords, resulting in the same $d_1$ but a larger $d_2$. For the same reason, any point on the ray $\bar{C}\bar{F}$ is also achievable (here we keep the same codebooks but decrease the rates for both users by the same amount). Any inner point of $\mathcal{D}_1^*$ can be expressed as the convex combination of two points, one from $\bar{C}\bar{H}$ and one from $\bar{C}\bar{F}$, and hence is also achievable due to Proposition \ref{convex}. The converse follows from the outer-bound provided in Lemma \ref{openregion} and the definition of $\mathcal{D}^*_1$ in Proposition \ref{convex} that $d_1\leq d_2$.

We now prove $\mathcal{D}_2^*=\{(d_1,d_2)|d_1\geq \tau_1/\gamma(P_1), d_2\geq \tau_2/\gamma(P_2), d_1\geq d_2,\gamma(P_1)d_1+[\gamma(P_1+P_2)-\gamma(P_1)]d_2\geq \tau_1+\tau_2\}$. For the achievability, notice that
\begin{align*}
    \bar{A}&=\mathbf{d}_2(A)=\left(\tfrac{\tau_1\gamma(P_2)+\tau_2(\gamma(P_1)+\gamma(P_2)-\gamma(P_1+P_2))} {\gamma(P_1)\gamma(P_2)},\tfrac{\tau_2}{\gamma(P_2)}\right),\\
    \bar{B}&=\mathbf{d}_2(B)=\left(\tfrac{\tau_1}{\gamma(P_1)},\tfrac{\tau_2}{\gamma(P_1+P_2)-\gamma(P_1)}\right)
\end{align*}
are achievable since they solve (\ref{optorigin}). Similar to the above argument, any point in $\mathcal{D}_2^*$ is achievable. For the converse, the first three inequalities defining $\mathcal{D}_2^*$ follow from the outer-bound in Lemma \ref{openregion} and the definition of $\mathcal{D}^*_2$ in Proposition \ref{convex}. Next we argue by contradiction that the fourth inequality (corresponding to the line connecting $\bar{A}$ and $\bar{B}$ in Fig. \ref{CTRegion} Case I) has to hold for any achievable completion time pair. Suppose there exits a point $\mathbf{d}'\in\mathcal{D}_2^*$ such that $\gamma(P_1)d_1'+[\gamma(P_1+P_2)-\gamma(P_1)]d_2'< \tau_1+\tau_2$. Hence for the weight $w_2=\frac{\gamma(P_1)}{\gamma(P_1+P_2)}$, $\mathbf{d}'$ results in a smaller weighted sum completion time than that of $\mathbf{d}_2(A)$. This contradicts with the fact that $\mathbf{d}_2(A)$ minimizes weighted sum completion time for the weight $w_2$ in Case I according to Theorem \ref{mini}.

The union of $\mathcal{D}_1^*$ and $\mathcal{D}_2^*$ gives us the expression of $\mathcal{D}^*$ shown in Case I.
\end{IEEEproof}

Referring to Fig. \ref{CTRegion} which depicts three possible shapes of $\mathcal{D}^*$ depending on the ratio of $\tau_1$ and $\tau_2$, as we can see, $\mathcal{D}^*$ is not convex in Case II. This can be verified by inspecting the slopes of line $\bar{A}\bar{C}$, $\bar{A}\bar{B}$ and $\bar{B}\bar{C}$, which are $-\frac{\gamma(P_1)}{\gamma(P_1+P_2)-\gamma(P_1)}$, $-\frac{\tau_1}{\tau_2}$, and $-\frac{\gamma(P_1+P_2)-\gamma(P_2)}{\gamma(P_2)}$ respectively and are sorted in ascending order according to the definition of Case II.

%%%%%%%%%%%%%%%%%%%%%%%%%%%%%%%%%%%%%%%%%%%%%%%%%%%%%%____Discussion_____%%%%%%%%%%%%%%%%%%%%%%%%%%%%%%%%%%%%%%%%%%%%%%%%%%%%%%%%
\section{Discussions}
Network design often incorporates the goal of optimizing a certain utility function, which for example can be a function of users' rates. Besides rate, another performance metric that is often of interest is delay. Equipped with the completion time region, one could optimize a utility that is a function of users' completion times. In this section, two particular optimizations are sought: minimization of the weighted sum completion time and minimization of the maximum completion time. The obtained solutions are information theoretically optimal in the sense that no reliable communication system can achieve a lower value.

First let us revisit the weighted sum completion time minimization problem (\ref{optorigin}) with the feasible set $\mathcal{D}_i$ being replaced by the whole completion time region $\mathcal{D}^*$, i.e. consider $\min_{(d_1,d_2)\in\mathcal{D}^*} wd_1+\bar{w}d_2$. The solution is given in the following theorem.
\begin{theorem}
\label{MACtheorem}
The minimum weighted sum completion time for the two-user GMAC defined in (\ref{GMACchannel}) is given by Table \ref{table2}, where $w_1,w_2$ are defined in Theorem \ref{mini}, $w_3=\frac{\tau_1}{\tau_1+\tau_2}$ and $D_i(\cdot)$ is defined in (\ref{Di}). Furthermore, the minimum value is attained at point $\mathbf{d}_i(\mathbf{r})$, where $\mathbf{r}$ is either $A$ or $B$ accordingly.
\begin{center}
\begin{table}[htbp]
\caption{}
\label{table2}
\hfill{}
\begin{tabular}{|c|c|c|}
\hline
     Case I  &  Case II &  Case III \\
\hline
$D_2(A)$, $w\in[0,w_2]$ & $D_2(A)$, $w\in[0,w_3]$ & $D_1(A)$, $w\in[0,w_1]$ \\\hline
$D_2(B)$, $w\in(w_2,1]$ & $D_1(B)$, $w\in(w_3,1]$ & $D_1(B)$, $w\in(w_1,1]$ \\
\hline
\end{tabular}
\hfill{}
\end{table}
\end{center}
\end{theorem}

\begin{IEEEproof}
Referring to Fig. \ref{CTRegion}, imagine there is a line with some fixed negative slope $s$ that moves towards the origin. When this line becomes tangent to $\mathcal{D}^*$, the tangent point will solve the weighted sum completion time minimization problem with the weight $w=\frac{s}{s-1}$. It is easy to see that the the problem is solved at either $A$ or $B$ depending on the weight. The weight, for which $A$ and $B$ result in equal weighted sum completion time, is determined by the slope of line $AB$ and is given by $w_i$ for each case. The detailed proof is technical and hence omitted.
\end{IEEEproof}

Next consider the maximum completion time minimization problem: $d_{m}=\min\max_{(d_1,d_2)\in\mathcal{D}^*}\{d_1,d_2\}$.
\begin{theorem}
\label{minimax}
The optimal value $d_{m}^*$ of the above minimax problem is given by the following table
\
\
\begin{center}
\begin{tabular}{|c|c|c|}
\hline
     Case I  &  Case II &  Case III \\
\hline
$\tfrac{\tau_1}{\gamma(P_1)}$ & $\frac{\tau_1+\tau_2}{\gamma(P_1+P_2)}$ & $\frac{\tau_2}{\gamma(P_2)}$ \\
\hline
\end{tabular}
\end{center}
\end{theorem}
\
\begin{IEEEproof}
Referring to Fig. \ref{CTRegion}, We will show the minimax problem is solved at point $\bar{C}$ for all cases. Hence the optimal value is given by the component of $\bar{C}$ (note that $\bar{C}$ has equal components). Consider $d_{m,1}=\min\max_{\mathbf{d}\in\mathcal{D}_1^*}\{d_1,d_2\} =\min_{\mathbf{d}\in\mathcal{D}_1^*} d_2$. Referring to Fig. \ref{CTRegion}, the minimum value $d_{m,1}^*$ is attained at point $\bar{C}$. Similarly for $d_{m,2}=\min\max_{\mathbf{d}\in\mathcal{D}_2^*}\{d_1,d_2\} =\min_{\mathbf{d}\in\mathcal{D}_2^*} d_1$, $d_{m,2}^*$ is also attained at $\bar{C}$. Hence $d_{m}^*=\min \{d_{m,1}^*,d_{m,2}^*\}$ is attained at $\bar{C}$.
\end{IEEEproof}

The information theoretic formulation of completion time discussed in this paper is based on the concept of constrained rates introduced through a two-user DMMAC. For the special case of GMAC, the explicit completion time region is then obtained. Since the Gaussianality of the channel does not come into the picture until the explicit formulation of the minimization problem in subsection III.B, all prior observations, particularly the convexity of sub-regions of the completion time region, also apply to a general MAC. Furthermore, it is not difficult to see that the concept of constrained rates is not unique to MAC and is applicable to other multi-user channels as well. Consequently, one line of further work is to study the completion time problem for other multi-user channels under the information theoretic framework introduced in this paper. This approach is pursued in \cite{Liu2}.

%%%%%%%%%%%%%%%%%%%%%%%%%%%%%%%%%%%%%%%%%%%%%%%____Appendix_____%%%%%%%%%%%%%%%%%%%%%%%%%%%%%%%%%%%%%%%%%%%%%%%%%%%%
\appendices
\section{Proof of Lemma \ref{raterelation}}
\label{Proofraterelation}
\begin{IEEEproof}
We prove for $c<1$, i.e. $n_1<n_2$. The case $c>1$ follows similarly. Consider time sharing between the coding scheme, which achieves $(R_1,R_2')\in\mathcal{C}_1$ in the first $n_1$ channel uses, and the coding scheme, which achieves rate $R_2''\leq R_{2}^*$ for user 2 in the remaining $n_2-n_1$ channel uses while user 1 transmits symbol $\phi_1$. Since in each sub-interval error probability can be made arbitrarily small, the overall time sharing scheme is reliable. User 2's overall rate therefore is given by $R_2=\log_2(M_2)/n_2=[n_1R_2'+(n_2-n_1)R_2'']/n_2=cR_2'+(1-c)R_2''$. Therefore $(R_1,R_2)$ is an achievable $c$-constrained rate pair. It is necessary to point out a likely false conclusion resulting from the time sharing argument: user 1's overall rate is $cR_1$, instead of $R_1$, because in the remaining $1-c$ fraction of time user 1 transmits at zero rate. This conclusion neglects the fact that user 1's rate $R_1=\log_2(M_1)/n_1$ is defined over $n_1$ (not $n_2$) channel uses.
\end{IEEEproof}

\section{Proof of Theorem \ref{maccapacity}}
\label{Proofmaccapacity}
\begin{IEEEproof}
We prove for $c<1$, i.e. $n_1< n_2$. The case $c>1$ follows similarly. The achievability follows from Lemma \ref{raterelation}. Specifically for $R_2\leq (1-c)R_2^*$, set $R_2'=0$ and $R_2''=\frac{1}{1-c}R_2$. For $R_2> (1-c)R_2^*$, set $R_2'=\frac{1}{c}R_2-(\frac{1}{c}-1)R_2^*$ and $R_2''=R_2^*$.

The converse is as the following. Let $Q$ denote a uniformly distributed r.v. on $\{1,...,n_1\}$. For an arbitrarily small $\epsilon$,
\begin{align}
    n_1R_1-n_1\epsilon &\leq I(X_1^{n_1};Y^{n_1}) \label{macr11}\\
    &\leq n_1I(X_1;Y|X_2,Q)\label{macr15}
%    n_1R_1&-n_1\epsilon\notag\\
%    &\leq I(W_1;Y^{n_1})\label{macr11}\\
%    &\leq I(X_1^{n_1},\phi_{1,n_1+1}^{n_2};Y^{n_1},Y_{n_1+1}^{n_2})\label{macr12}\\
%    %&= I(X_1^{n_1};Y^{n_1})+I(X_1^{n_1};Y_{n_1+1}^{n_2}|Y^{n_1})\label{macr13}\\
%    &= I(X_1^{n_1};Y_{n_1+1}^{n_2})+I(X_1^{n_1};Y^{n_1}|Y_{n_1+1}^{n_2})\label{macr13}\\
%    &= H(X_1^{n_1})-H(X_1^{n_1}|Y^{n_1},Y_{n_1+1}^{n_2})\label{macr1new1}\\
%    &\leq H(X_1^{n_1}|X_2^{n_1})-H(X_1^{n_1}|Y^{n_1},Y_{n_1+1}^{n_2},X_{2}^{n_2})\label{macr1new2}\\
%    &= I(X_1^{n_1};Y^{n_1}|X_2^{n_1})\label{macr1new3}\\
%    &\leq n_1I(X_1;Y|X_2,Q)\label{macr15},
\end{align}
where (\ref{macr11}) is due to Fano's inequality and data processing inequality. Detailed steps from (\ref{macr11}) to (\ref{macr15}) can be found in equations 15.104-15.113 in section 15.3.4 \cite{Cover}.

Similarly, due to Fano's inequality and data processing inequality, we have
\begin{align}
    &n_2(R_2-\epsilon)\notag\\
    &\leq I(X_2^{n_2};Y^{n_2})\notag\\
    &\leq I(X_2^{n_1},X_{2,n_1+1}^{n_2};Y^{n_1},Y_{n_1+1}^{n_2}|X_1^{n_1})\label{macr2new1}\\
    &=I(X_2^{n_1};Y^{n_1}|X_1^{n_1})+I(X_2^{n_1};Y_{n_1+1}^{n_2}|X_1^{n_1},Y^{n_1})\notag\\
    &\quad +I(X_{2,n_1+1}^{n_2};Y_{n_1+1}^{n_2}|X_1^{n_1},X_2^{n_1}) \notag\\
    &\quad +I(X_{2,n_1+1}^{n_2};Y^{n_1}|X_1^{n_1},X_2^{n_1},Y_{n_1+1}^{n_2})\notag\\
    &= I(X_2^{n_1};Y^{n_1}|X_1^{n_1})+I(X_2^{n_1};Y_{n_1+1}^{n_2}|X_1^{n_1},Y^{n_1})\notag\\
    &\quad +I(X_{2,n_1+1}^{n_2};Y_{n_1+1}^{n_2}|X_1^{n_1},X_2^{n_1},Y^{n_1}) \label{macr2new2}\\
    &= I(X_2^{n_1};Y^{n_1}|X_1^{n_1})+ I(X_2^{n_1},X_{2,n_1+1}^{n_2};Y_{n_1+1}^{n_2}|X_1^{n_1},Y^{n_1})\notag\\
    &= I(X_2^{n_1};Y^{n_1}|X_1^{n_1})+ I(X_{2,n_1+1}^{n_2};Y_{n_1+1}^{n_2}|X_1^{n_1},Y^{n_1})\notag\\
    &\quad +I(X_2^{n_1};Y_{n_1+1}^{n_2}|X_1^{n_1},Y^{n_1},X_{2,n_1+1}^{n_2})\notag\\
    &= I(X_2^{n_1};Y^{n_1}|X_1^{n_1})+ I(X_{2,n_1+1}^{n_2};Y_{n_1+1}^{n_2}|X_1^{n_1},Y^{n_1})\label{macr2new3}\\
    &\leq I(X_2^{n_1};Y^{n_1}|X_1^{n_1})+I(X_{2,n_1+1}^{n_2};Y_{n_1+1}^{n_2}|\phi_{1,n_1+1}^{n_2})\label{macr21}
\end{align}
where (\ref{macr2new1}) is due to conditioning reduces entropy and $X_1$ and $X_2$ are independent, (\ref{macr2new2}) is because $Y^{n_1}$ is independent of others conditioned on $X_1^{n_1}$ and $X_2^{n_1}$, (\ref{macr2new3}) is because $Y_{n_1+1}^{n_2}$ is independent of others conditioned on $X_{2,n_1+1}^{n_2}$, (\ref{macr21}) is due to conditioning reduces entropy, $Y_{n_1+1}^{n_2}$ is independent of others conditioned on $X_{2,n_1+1}^{n_2}$ and the fact that $\phi_1$ is a constant. Using the argument in section 15.3.4 \cite{Cover}, it can be shown $I(X_2^{n_1};Y^{n_1}|X_1^{n_1})\leq n_1 I(X_2;Y|X_1,Q)$. Considering a DMC with channel transition probability $p(y|x_1=\phi_1,x_2)$, we have $I(X_{2,n_1+1}^{n_2};Y_{n_1+1}^{n_2}|\phi_{1,n_1+1}^{n_2})\leq (n_2-n_1)R_2^*$, where $R_2^*$ is defined in (\ref{ratestar}). Therefore we have
\begin{equation}
    \tfrac{1}{c}R_2-(\tfrac{1}{c}-1)R_2^*-\tfrac{1}{c}\epsilon\leq I(X_2;Y|X_1,Q). \label{macr22}
\end{equation}
Again due to Fano's inequality and data processing inequality, we have the sum rate upper-bound
\begin{align}
    &n_1R_1+n_2R_2-n_2\epsilon\notag\\
    &\leq I(X_1^{n_1},X_2^{n_1},X_{2,n_1+1}^{n_2};Y^{n_1},Y_{n_1+1}^{n_2})\notag\\
    &= I(X_1^{n_1},X_2^{n_1};Y^{n_1})+I(X_1^{n_1},X_2^{n_1};Y_{n_1+1}^{n_2}|Y^{n_1})\notag\\
    &\quad +I(X_{2,n_1+1}^{n_2};Y_{n_1+1}^{n_2}|X_1^{n_1},X_2^{n_1})\notag\\
    &\quad +I(X_{2,n_1+1}^{n_2};Y^{n_1}|X_1^{n_1},X_2^{n_1},Y_{n_1+1}^{n_2})\notag
\end{align}
\begin{align}
    &= I(X_1^{n_1},X_2^{n_1};Y^{n_1})+I(X_1^{n_1},X_2^{n_1};Y_{n_1+1}^{n_2}|Y^{n_1})\notag\\
    &\quad +I(X_{2,n_1+1}^{n_2};Y_{n_1+1}^{n_2}|X_1^{n_1},X_2^{n_1},Y^{n_1})\label{macrsnew1}\\
    &= I(X_1^{n_1},X_2^{n_1};Y^{n_1}) + I(X_1^{n_1},X_2^{n_1},X_{2,n_1+1}^{n_2};Y_{n_1+1}^{n_2}|Y^{n_1})\notag\\
    &= I(X_1^{n_1},X_2^{n_1};Y^{n_1}) + I(X_{2,n_1+1}^{n_2};Y_{n_1+1}^{n_2}|Y^{n_1})\notag\\
    &\quad +I(X_1^{n_1},X_2^{n_1};Y_{n_1+1}^{n_2}|Y^{n_1},X_{2,n_1+1}^{n_2})\notag\\
    &\leq I(X_1^{n_1},X_2^{n_1};Y^{n_1})+I(X_{2,n_1+1}^{n_2};Y_{n_1+1}^{n_2}|\phi_{1,n_1+1}^{n_2})\label{macrs1}\\
    &\leq n_1I(X_1,X_2;Y|Q)+(n_2-n_1)R_2^*\notag,
\end{align}
where (\ref{macrsnew1}) is because $Y^{n_1}$ is independent of others conditioned on $X_1^{n_1},X_2^{n_1}$, (\ref{macrs1}) is due to conditioning reduces entropy, $Y_{n_1+1}^{n_2}$ is independent of others conditioned on $X_{2,n_1+1}^{n_2}$ and the fact that $\phi_1$ is a constant. Therefore we have
\begin{equation}
    R_1+\tfrac{1}{c}R_2-(\tfrac{1}{c}-1)R_2^*-\tfrac{1}{c}\epsilon\leq I(X_1,X_2;Y|Q).\label{macrss}
\end{equation}

Note that the mutual information terms on line (\ref{macr15}), (\ref{macr22}) and (\ref{macrss}) altogether define the standard MAC capacity region.
\end{IEEEproof}

\section{}
\label{Proofmini}

Due to the equivalency of the two problems, to solve (\ref{optorigin}), we consider (\ref{newopt}). Note that (\ref{newopt}) is a non-convex optimization problem since the objective function is not convex. Fortunately, using the following Lemma, we can obtain a closed form solution.

\begin{definition}
Let $\mathbb{R}_c^2$ be a convex subset of $\mathbb{R}^2$. A point $\mathbf{r}_x\in\mathbb{R}_c^2$ is an \textit{extreme point} iff whenever $\mathbf{r}_x=t\mathbf{r}_y+(1-t)\mathbf{r}_z$, $t\in(0,1)$ and $\mathbf{r}_y\neq \mathbf{r}_z$, this implies either $\mathbf{r}_y\not \in \mathbb{R}_c^2$ or $\mathbf{r}_z\not \in \mathbb{R}_c^2$.
\end{definition}
\begin{definition}
An extreme point $\mathbf{r}_x\in\mathbb{R}_c^2$ is said to be \textit{dominant} iff there does not exist extreme point $\mathbf{r}_y\in\mathbb{R}_c^2$, $\mathbf{r}_y\neq \mathbf{r}_x$, such that $\mathbf{r}_x\leq \mathbf{r}_y$ element-wise.
\end{definition}
\begin{lemma}
\label{expoint}
The minimum value of (\ref{newopt}) is attained at dominant extreme points of $\mathcal{C}_{1,i}^G$.
\end{lemma}
\begin{IEEEproof}
Assume $i=1$. The case $i=2$ can be obtained similarly. Suppose $\mathbf{r}_x\in\mathcal{C}_{1,1}^G$ is not an extreme point, i.e. there exist points $\mathbf{r}_y,\mathbf{r}_z\in\mathcal{C}_{1,1}^G$,  and $\mathbf{r}_y,\mathbf{r}_z\neq \mathbf{r}_x$ such that $\mathbf{r}_x$ lies on the line segment, of which the end points are $\mathbf{r}_y$ and $\mathbf{r}_z$. Let this line segment be represented by $ar_1+br_2=1$. Next we evaluate $D_1$ (\ref{Di}) along this line segment.
\begin{enumerate}
\item For $b\neq 0$, without loss of generality, suppose $r_{1,y}< r_{1,x}< r_{1,z}$. We have $r_2=\frac{1-ar_1}{b}$ and thus $D_1=\bar{w}\frac{\tau_2}{R_2^*} +\frac{\tau_1}{R_2^*}\left(\frac{R_2^*-\frac{1}{b}\bar{w}}{r_1}+\bar{w}\frac{a}{b}\right)$, which is an increasing function of $r_1$ if $R_2^*-\frac{1}{b}\bar{w}<0$ or decreasing if $R_2^*-\frac{1}{b}\bar{w}>0$. Therefore we have $\min\{D_1(\mathbf{r}_x), D_1(\mathbf{r}_y),D_1(\mathbf{r}_z)\}\neq D_1(\mathbf{r}_x)$ unless $R_2^*-\frac{1}{b}\bar{w}=0$, in which case $D_1(\mathbf{r}_x)=D_1(\mathbf{r}_y)=D_1(\mathbf{r}_z)$.
\item For $b=0$, then $r_{1,y}=r_{1,x}=r_{1,z}=\frac{1}{a}$ and $D_1=\bar{w}\frac{\tau_2}{R_2^*}+\frac{a\tau_1(R_2^*-\bar{w}r_2)}{R_2^*}$ which is a decreasing function of $r_2$. Without loss of generality we assume $r_{2,y}< r_{2,x} < r_{2,z}$. Then we have $D_1(\mathbf{r}_x)> D_1(\mathbf{r}_z)$.
\end{enumerate}
In either case, $D_1(\mathbf{r}_x)\geq \min\{D_1(\mathbf{r}_y),D_1(\mathbf{r}_z)\}$. Hence it is sufficient to consider extreme points only. Furthermore, since $D_1$ is a decreasing function of $r_1$ and $r_2$, we need only focus on dominant extreme points.
\end{IEEEproof}

We are now in position to solve (\ref{newopt}).
\begin{IEEEproof}[Proof of Theorem \ref{mini}]
We refer (\ref{newopt}) with $i=1$ ($i=2$) as problem 1 (2). We solve problem 1. Problem 2 follows similarly.

\noindent \textit{1. Case I}

Referring to Case I shown in Fig \ref{MAC capacity region}, the feasible set $\mathcal{C}_{1,1}^G$ is the triangle OEC, which has only one dominant extreme point $C$. By Lemma \ref{expoint}, problem 1 is solved at point $C$ for all $w$.

\noindent \textit{2. Case II}

Referring to Case II shown in Fig \ref{MAC capacity region}, $\mathcal{C}_{1,1}^G$ is the quadrangle OEBC, which has two dominant extreme points $B,C$. By Lemma \ref{expoint}, $D_1^*=\min\{D_1(B),D_1(C)\}$. Since $B,C$ are both on the line $r_1+r_2=\gamma(P_1+P_2)$, in order to compare $D_1(B)$ and $D_1(C)$, we plug the line equation and $R_2^*=\gamma(P_2)$ into $D_1$ in (\ref{Di}) and obtain $D_1=\bar{w}\frac{\tau_2}{\gamma(P_2)}+\frac{\tau_1}{\gamma(P_2)}[\frac{\gamma(P_2)-\bar{w}\gamma(P_1+P_2)}{r_1}+\bar{w}]$. Hence we have $D_1^*=D_1(C)$ for $w\in[0,w_1]$, i.e. problem 1 is solved at point $C$ for $w\in[0,w_1]$. Similarly problem 1 is solved at point $B$ for $w\in(w_1,1]$.

\noindent \textit{3. Case III}

Following the similar argument in Case II, we can show that problem 1 is solved at point $A$ for $w\in[0,w_1]$ and point $B$ for $w\in[0,w_1]$.

Problem (\ref{newopt}) is solved at one of the rate points $A$, $B$ and $C$ according to Table \ref{table}. Therefore, by Proposition \ref{eq}, the completion time pairs shown in Table \ref{table} solve problem (\ref{optorigin}).

\end{IEEEproof}


\begin{thebibliography}{1}

\bibitem{Ahlswede}
R. Ahlswede, ``Multi-way communication channels,'' \textit{IEEE Int. Symp. Inf. Theory}, Tsahkadsor, Armenian S.S.R., 1971.

\bibitem{Liao}
H. Liao, ``Coding theorem for multiple access communications,'' \textit{Int. Symp. Inf. Theoy}, Asilomar, CA, 1972.

\bibitem{Abramson}
N. Abramson, ``The ALOHA system - Another alternative for computer communications,'' in \textit{Proc. Full Joint Computer Conf.}, AFIPS Conf., vol. 37, 1970, pp. 281-285.

\bibitem{Gallager}
R. G. Gallager, `` A perspective on multiaccess channels,'' \textit{IEEE Trans. Inf. Theory}, vol. 31, pp. 124-142, Mar. 1985.

\bibitem{Ephremides}
A. Ephremides and B. Hajek, ``Information theory and communication networks: an unconsummated union,'' \textit{IEEE Trans. Inf. Theory}, vol. 44, pp. 2416-434, Oct 1998.

\bibitem{Rai}
S. Raj, E. Telatar, and D. Tse, ``Job scheduling and multiple access,'' DIMACS Series in Discrete Mathematics and Theoretical Computer Sciences, vol. 66, pp. 127-37, 2003.

\bibitem{Ng}
C.T.K. Ng, M. Medard and A. Ozdaglar, ``Completion time minimization and robust power control in wireless packet networks,'' available on http://arxiv.org/abs/0812.3447, Apr 2011.

\bibitem{Cover}
T. Cover and J. Thomas, \textit{Elements of Information Theory 2nd Edt}, New York: Wiley, 2006.

\bibitem{Liu2}
Y. Liu and E. Erkip, ``Completion time in broadcast channel and interference channel,'' to appear, \textit{Proc. Allerton Conference on CCC}, Sep. 2011.

\end{thebibliography}
\end{document}